\documentclass[prc,fleqn,twoside,floatfix]{revtex4}
\usepackage{latexsym,exscale}
\usepackage{epsfig}
\usepackage{amsmath,amssymb,amsfonts}
\begin{document}
\title{Sum rules and short-range correlations in
nuclear matter at finite temperature}
\author{T. Frick and H. M\"uther}
\affiliation{Institut f\"ur
Theoretische Physik, \\ Universit\"at T\"ubingen, D-72076 T\"ubingen, Germany}
\author{ A. Polls}
\affiliation{Departament d'Estructura i Constituents de la Mat\`eria,\\
Universitat de Barcelona, E-08028 Barcelona, Spain}

\begin{abstract}
The nucleon spectral function in nuclear matter fulfills an energy
weighted sum rule. Comparing two different realistic potential, these sum
rules are studied for Green's functions that are derived
self-consistently within the $T$~matrix approximation at finite
temperature.   
\end{abstract}
%\pacs{21.65.+f, 21.30.Fe}
\maketitle

%\section{Introduction\label{Introduction}}      

The microscopic study of the single-particle properties in nuclear matter
requires a rigorous treatment of the nucleon-nucleon (NN) correlations
\cite{bal1,her1}.
In fact, the strong short-range and tensor components, which are needed
in realistic NN interactions to fit the NN scattering data, lead to 
important modifications of the nuclear wave function. A clear indication of
the importance of correlations is provided by the observation that a simple 
Hartree-Fock calculation for nuclear matter at the empirical
saturation 
density using
such realistic NN interactions typically results in positive energies
rather 
than the 
empirical value of $-16\,\mbox{MeV}$ per nucleon \cite{her1}. 

Correlations do not only manifest themselves
in the bulk properties but also modify the single-particle properties in a
substantial way. Several recent calculations have shown without
ambiguity 
how the 
NN correlations produce a partial occupation of the single particle 
states which would
be fully occupied in a mean field description and a wide distribution 
in energy of
the single particle strength. These two features have also been
empirically 
founded
in the analysis of the (e,e'p) nucleon knock-out reactions \cite{exp1}. 
The theoretical studies have been conducted both in finite 
nuclei \cite{her2} and also in
nuclear matter \cite{fan1,ram1,dick1}.

An optimal tool to study the single particle properties is provided 
by the self-consistent
Green's function technique (SCGF) \cite{dick2}. 
This method gives direct access to the single particle 
spectral function, which should be self-consistently determined at the same time
than the effective interactions between the nucleons in the medium.  
Enormous progress
in the SCGF applications to nuclear matter have been reported in the 
last years, both at zero
\cite{dick1}  and finite temperature \cite{boz1,boz2,fri03}.

The efforts at $T=0$ have mainly been addressed to provide the
appropriate theoretical
background for the interpretation of the (e,e'p) experiments while the 
investigation
at finite $T$ are mainly oriented to describe the nuclear medium in astrophysical
environments or to the interpretation of the dynamics of heavy ion  collisions.

In any case, the key quantity is the single-particle spectral
function, 
i.e. the distribution
of strength in energy  when one adds or removes a particle of the
system 
with a given
momentum. A possible way to analyze the single-particle spectral
function is by means 
of the energy weighted sum rules. They are well established in the
literature and have been numerically analyzed in the case of zero
temperature \cite{pol94}.  

The analysis of the energy weighted sum-rules can give useful 
insights not only on the 
numerical accuracy of the many-body approach used to calculate them
but also can help to 
understand the properties and structure of the NN potential.

This paper is devoted to study the physical implications of the
fulfillment of these
sum rules for single particle spectral functions in nuclear matter at
finite $T$. This investigation is based on the
framework of SCGF employing a fully self-consistent ladder approximation in
which the complete
spectral function has been used to describe the intermediate states in
the Galistkii-Feynman
equation.

After a brief summary of the definitions of the single-particle
spectral function,  
we give a simple derivation of the sum rules. Then we analyze the
results for two types 
of realistic potentials,  the CDBONN and the Argonne V18, and discuss
the different behaviors 
based on the different strength of the short-range and tensor
components of both potentials. 

%\section{Sum rules}

For a given Hamiltonian $H$, 
the Green's function for a system at finite temperature can 
be defined in a grand-canonical formulation:
\begin{equation}
\label{green_def}
{\mathrm{i}}g({\mathbf{k}}t;{\mathbf{k}}^{\prime}t^{\prime})=
{\mathrm{Tr}}\{\rho\,
%{\mathbb{T}}
{\mathbf{T}}
[a^{\phantom{\dagger}}_{\mathbf{k}}(t)
a^{\dagger}_{{\mathbf{k}}^{\prime}}(t^{\prime})]\}. 
\end{equation}
$\mathbf{T}$ is the time ordering operator that acts on a product
of Heisenberg field operators
$a_{\mathbf{k}}(t)=e^{{\mathrm{i}}tH}a_{\mathbf{k}}e^{-{\mathrm{i}}tH}$
in such a way that the field operator with the largest time argument $t$
is put to the left.
The trace is to be taken over all energy eigenstates and all particle number
eigenstates of the many body system, weighted by the statistical operator,
\begin{equation}
\label{eq_statop}
\rho=\frac{1}{Z}\,{e^{-\beta(H-\mu N)}}.
\end{equation}
$\beta$ and $\mu$ denote the inverse temperature and the chemical 
potential, respectively. 
N is the operator that counts the total number of
particles in the system,
\begin{equation}
N=
\sum_{\mathbf{k}}
a_{\mathbf{k}}^{\dagger}(t)a_{\mathbf{k}}^{\phantom{\dagger}}(t)
%\int\!{\mathrm{d}}^3\!x\,
%\psi^{\dagger}({\mathbf{x}t})
%\psi({\mathbf{x}t}).
\end{equation}
$N$ is independent of time, since it commutes with $H$.
The normalization factor in Eq.~(\ref{eq_statop}) is given by the
grand partition function of statistical mechanics,
\begin{equation}
Z={{\mathrm{Tr}}\,e^{-\beta(H-\mu N)}}.
\end{equation}
For a homogeneous system, the Green's function is diagonal in 
momentum space and depends only on the absolute value of 
${\mathbf{k}}$ and on the difference $\tau=t^{\prime}-t$.
Starting from the definition of the Green's function, 
we first focus on the case $\tau>0$.
In order to recover the 
expression for the ensemble average of the occupation number 
$n(k)$ for $\tau=0^+$,
the following definition of the correlation function, $g^<$, 
includes an additional factor of $-{\mathrm{i}}$ with respect to the  
definition of the Green's function $g$, 
\begin{equation}
g^<(k,\tau)=
\label{eq_gs1}
{\mathrm{Tr}}\{\rho\,
e^{{\mathrm{i}}\tau H}a^{\dagger}_{{\mathbf{k}}}
e^{-{\mathrm{i}}\tau H}
a^{\phantom{\dagger}}_{\mathbf{k}}\}.
\end{equation} 
$g^<(k,\tau)$ can be expressed as a Fourier integral over all frequencies, 
\begin{equation}
\label{eq_momdist}
g^<(k,\tau)=\int_{-\infty}^{+\infty}
\frac{{\mathrm{d}}\omega}{2\pi}\,e^{-{\mathrm{i}}\omega \tau}A^<(k,\omega)
\end{equation}
if $A^<(k,\omega)$ is defined by~\cite{kraeft}:
\begin{equation}
\label{eq_gsmaller}
A^<(k,\omega)=2\pi 
%\frac{1}{Z}
\sum_{nm}
\frac{e^{-\beta(E_m-\mu N_m)}}{Z}
\left|\left<\Psi_n|a_k|\Psi_m\right>\right|^2
\delta(\omega-(E_m-E_n)).
\end{equation}
This can be easily checked by inserting the eigenstates
$\left|\Psi_n\right>$ into the expression of the trace in 
Eq.~(\ref{eq_gs1}). It is important to 
note that $\left|\Psi_n\right>$ are simultaneous
eigenstates of both the number operator and the Hamiltonian.\\  
A similar analysis can be conducted for $\tau<0$, yielding a 
function 
\begin{equation}
A^>(k,\omega)=e^{\beta(\omega-\mu)}A^<(k,\omega).\label{eq:connec}
\end{equation} 
The spectral function at finite temperature is defined as the sum 
of the two positive functions, $A^<$ and $A^>$,
\begin{equation}
A(k,\omega)=A^<(k,\omega)+A^>(k,\omega).
\end{equation} 
Expression~(\ref{eq_gsmaller}), for $A^<$, can be compared
to the result for the hole spectral function 
at zero temperature, that was reported in Ref~\cite{pol94},
\begin{equation}
A_h(k,\omega)= 2\pi 
\sum_{n}
\left|\left<\Psi_n^{A-1}|a_k|\Psi_0^{A}\right>\right|^2
\delta(\omega-(E_0^{A}-E_n^{A-1})),
\end{equation}
where 
$\left|\Psi_0^{A}\right>$ is the ground state of an $A$ particle system and 
$\left|\Psi_n^{A-1}\right>$ labels the excited energy eigenstates of a system 
that contains one particle less. 
The physical interpretation of the hole spectral function in a system 
at zero temperature is the following: $A_h(k,\omega)$ is the probability 
to remove a particle from the ground state of the $A$-body system, such, 
that the residual system
is left with an excitation energy $E^{A-1}_n=E^A_0-\omega$. 
$E^A_0$ is the ground state energy of the $A$ particle system.
It is clear that the lowest possible energy of the final state is 
the ground state energy of the $A-1$ particle system, so that there 
is an upper limit for the hole spectral function at 
$\omega=E_0^A-E_0^{A-1}=\mu$. In a similar fashion, the particle spectral 
function $A_p$ can be defined as the probability to attach a further nucleon 
to the system in such a way, that the excitation energy of the compound 
system with respect to the ground state energy of the initial system is 
$\omega=E^{A+1}-E_0^{A}$. In this case, one can argue that, 
to add a further particle, one has to pay 
at least the chemical potential, so that $\mu$ is a lower bound for 
$\omega$. At zero temperature, 
this behavior causes a complete separation of the particle and the 
hole spectral function. 

\begin{figure}
\begin{center}
\epsfig{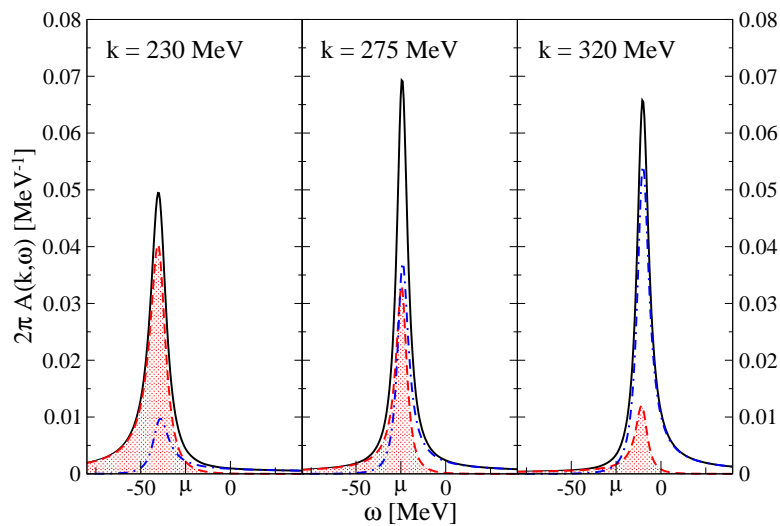}
\end{center}
\caption{\label{fig_sf}Spectral function for a density of 
$\rho=0.2\,\mbox{fm}^{-3}$ and a temperature of $T=10\,\mbox{MeV}$ 
(solid line). Various momenta are considered as indicated in the three panels.
$A^<$ (dashed line) and $A^>$ (dash-dotted line) are also displayed.}
\end{figure}

The situation is quite different in a grand-canonical formulation at 
finite temperature 
To illustrate these changes, the full spectral function A,
%The spectral function, 
as well as $A^>$ and $A^<$ 
%at finite temperature is 
are shown in Fig.~\ref{fig_sf} for three momenta around the Fermi momentum 
of a zero temperature system at the same density, $\rho=0.2\,\mbox{fm}^{-3}$.
Numerical values for the integrated strength of $A^<$ are listed in Table 
\ref{tab_dist}. 
Since thermally excited states $\left|\Psi_m\right>$ 
are always included in the grand-canonical ensemble 
average according to their weight factor $e^{-\beta(E_m-\mu N_m)}$, 
one can take out a particle from a thermally excited 
state and end up in a weakly excited state close to 
the ground state of the residual system. This leads to a contribution to $A^<$
for an energy $\omega$ larger than $\mu$. Also one has to keep in mind that we
are considering a grand-canonical average. This implies that with the
appropriate weight one also considers systems with a density larger than the
mean value. For those systems,  a removal of 
particles from states with single-particle energies
above $\mu$ will also be possible.
 
Similarly, a particle can be added to a thermally excited state, 
leaving the compound system in a state close to its ground state, 
so that $A^>(k,\omega)$ 
extends to the region below $\mu$. 
In any case, there is no longer a separation between $A^>$ and $A^<$, 
and the maxima of both functions can even coincide. This is also quite obvious
from the relation (\ref{eq:connec}).

\begin{figure}
\begin{center}
\epsfig{figure=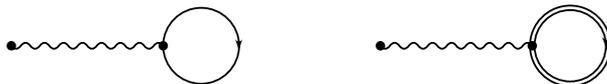,width=8cm}
\end{center}
\caption{\label{fig_tadpole}Diagrammatic representation of the HF
approximation (left) and the energy independent part of the
self-consistently dressed self energy (right).}
\end{figure}

For the $T$~matrix approximation to the self energy 
reported in~\cite{fri03}, one can determine the 
single-particle Green's function as the solution of
Dyson's equation for any complex value of the frequency variable $z$,
\begin{equation}
\label{dyson_g}
g(k,z)=
\frac{1}{z-\frac{k^2}{2m}-\Sigma(k,z)}.
\end{equation}
Using the analytical properties of the finite temperature Green's 
function along the imaginary time axis, an important relation 
between the spectral function and the Green's function can be 
derived and analytically continued to slightly complex values~\cite{kad62}:
%Using the periodicity of the finite temperature Green's function along
%the real-time axis, the following relation between $g$ and $A$ can be
%derived~\cite{kad62},  
\begin{equation}
\label{spec_g}
%g(k,z_{\nu})=\int_{-\infty}^{+\infty}
%\frac{{\mathrm{d}}\omega}{2\pi}\, \frac{A(k,\omega)}{z_{\nu}-\omega}.
g(k,\omega+{\mathrm{i}}\eta)=\int_{-\infty}^{+\infty}
\frac{{\mathrm{d}}\omega^{\prime}}{2\pi}\, \frac{A(k,\omega)}
{\omega-\omega^{\prime}+{\mathrm{i}}\eta}.
\end{equation}
One can extract sum rules from the asymptotic
behavior at large $\omega$ 
by expanding the real part of both expressions for the Green's function, 
Eqs.~(\ref{dyson_g}) and (\ref{spec_g}), in powers of
$\frac{1}{\omega}$. This yields
\begin{equation}
{\mathrm{Re}}\,g(k,\omega)=\frac{1}{\omega}
%{\mathrm{Re}}\,g(k,\omega+{\mathrm{i}}\eta)=\frac{1}{\omega}
\left\{1+\frac{1}{\omega}\left[
%\frac{k^2}{2m}+\Sigma^{\infty}(k)
\frac{k^2}{2m}+\lim_{\omega\rightarrow\infty}{\mathrm{Re}}\,
\Sigma(k,\omega)
%\Sigma(k,\omega+{\mathrm{i}}\eta)
\right]+\cdots\right\}
\end{equation} 
and 
\begin{equation}
{\mathrm{Re}}\,g(k,\omega)=
%{\mathrm{Re}}\,g(k,\omega+{\mathrm{i}}\eta)=
%\frac{1}{\omega}
\frac{1}{\omega}
\left\{
\int_{-\infty}^{+\infty}
{\mathrm{d}}\omega^{\prime}\,
A(k,\omega^{\prime})+
\frac{1}{\omega}
\int_{-\infty}^{+\infty}
{\mathrm{d}}\omega^{\prime}\,
\omega^{\prime}A(k,\omega^{\prime})+\cdots\right\}.
\end{equation}
By comparing the first two expansion coefficients, one finds the $m_0$
and the $m_1$ sum rules,
\begin{equation}
\int_{-\infty}^{+\infty} \frac{{\mathrm{d}}\omega}{2\pi}
A(k,\omega)=1,
\end{equation}
and
\begin{equation}
\int_{-\infty}^{+\infty} \frac{{\mathrm{d}}\omega}{2\pi}
A(k,\omega)\omega=\frac{k^2}{2m}+\lim_{\omega\rightarrow\infty}{\mathrm{Re}}\,
\Sigma(k,\omega)
\end{equation}
Similar sum rules can be obtained from the higher order terms,
as it was done in Ref.~\cite{pol94} for $m_2$. 
Thinking of an arbitrary
approximation scheme for $\Sigma(k,\omega)$, it might be
interesting to ask whether or to what extend such a scheme fulfills
the sum rules. This is, however, not the point we want to address in
this paper. In the $T$~matrix approximation, the real part of
the self energy can be computed from the imaginary part, using a
dispersion relation,
\begin{equation}
\label{eq_real_sigma}
{\mathrm{Re}}\,\Sigma(k,\omega)=\Sigma^{\infty}(k)-\frac{{\mathcal{P}}}{\pi}
\int_{-\infty}^{+\infty}
{\mathrm{d}}\lambda\,
\frac{{\mathrm{Im}}\Sigma(k,\lambda+{\mathrm{i}}\eta)}{\omega-\lambda}.
%+{\mathrm{i}}\,{\mathrm{Im}}\Sigma(k,\omega+{\mathrm{i}}\eta),
\end{equation}
In the derivation of Eq.~(\ref{eq_real_sigma}), the spectral
decomposition of the Green's function was already used, so it is a
property of the $T$~matrix approach that it automatically fulfills the sum
rules. Nevertheless, besides providing a useful consistency
check for the numerics, it is interesting to use the sum rules 
to compare the importance of
short-range correlations for different realistic 
potentials on a quantitative level.
The first term on the right hand side of Eq.~(\ref{eq_real_sigma}) is the energy independent part
of the self energy, 
\begin{equation}
\label{eq_HF}
\Sigma^{\infty}(k)
=
\int \frac{{\mathrm{d}}^3k^{\prime}}{(2\pi)^3}
\left<{\mathbf{k}}{\mathbf{k}}^{\prime}\right|
V
\left|{\mathbf{k}}{\mathbf{k}}^{\prime}\right>_A
n(k^{\prime}).
\end{equation}
%\begin{equation}
%\label{eq_HF}
%\Sigma^{\infty}(k)=
%\frac{1}{\beta}
%\sum_{\nu}
%\int \frac{{\mathrm{d}}^3k^{\prime}}{(2\pi)^3}
%\left<{\mathbf{k}}{\mathbf{k}}^{\prime}\right|V
%\left|{\mathbf{k}}{\mathbf{k}}^{\prime}\right>_A
%%\left<{\mathbf{k}}{\mathbf{k}}^{\prime}\right|V
%%\left|{\mathbf{k}}^{\prime}{\mathbf{k}}\right>
%%\right]
%g(k^{\prime},z_{\nu}),
%\end{equation}
which can be identified with $\lim_{\omega\rightarrow\infty}{\mathrm{Re}}\,
\Sigma(k,\omega)$, since the dispersive part decays like
$\frac{1}{\omega}$ for $\omega\rightarrow\pm\infty$.
Eq.~(\ref{eq_HF}) looks like a Hartree-Fock potential, however, $n(k)$ is 
the momentum distribution that is determined from a non-trivial 
spectral function $A^<$ in Eq.~(\ref{eq_momdist}), assuming $\tau=0$.
%\begin{equation}
%n(k)=
%\int_{-\infty}^{+\infty} \frac{{\mathrm{d}}\omega}{2\pi}
%A^<(k,\omega)
%\end{equation}
In contrast, the Hartree-Fock self energy at finite temperature 
must be determined from an energy spectrum 
$\epsilon(k)$ and a momentum distribution 
$n_{HF}(k)=f(\epsilon(k))$, where $f(\omega)$ is the Fermi function.
Unlike $n_{HF}(k)$, the non-trivial $n(k)$ accounts for depletion effects of 
the bound states due to short-range correlations.
In this sense, $\Sigma^{\infty}$ is a generalization of a Hartree-Fock
potential. Fig.~\ref{fig_tadpole} illustrates the difference between the
two pictures with the corresponding Feynman diagrams.

%\section{Results and discussion}

All results in this paper have been obtained using the
iteration procedure that was described in 
Ref.\cite{fri03}. Fully self-consistent spectral functions were 
calculated for
%at a density of $\rho=0.2\,\mbox{fm}^{-3}$. 
two realistic potentials, 
the stiffer Argonne V18 and the softer CDBONN.
%, have been used as an input. 

\begin{figure}
\begin{center}
\epsfig{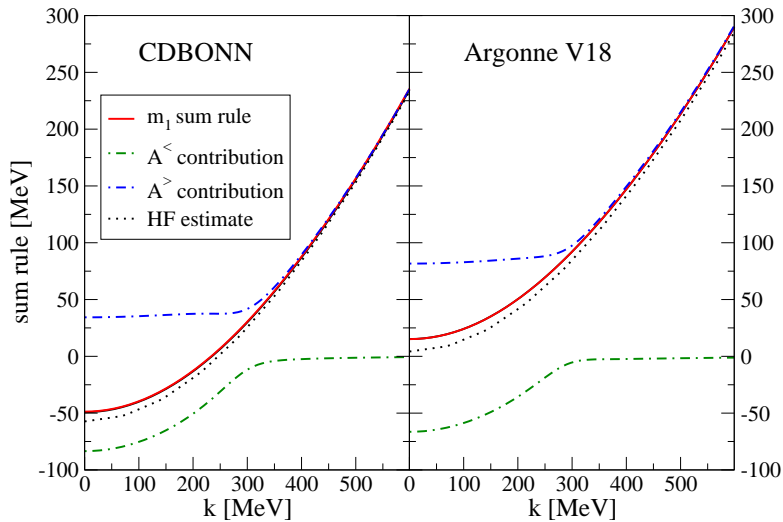}
\end{center}
\caption{\label{fig_m1}Illustration of the energy weighted sum rule 
$m_1$ (solid lines) for the CDBONN potential (left panel) 
and the Argonne V18 potential (right panel). 
Both right hand side and left hand side are displayed, 
but the sum rule is so well fulfilled that they are on top of each other. 
The contribution to $m_1$ that comes from $A^>$ 
and $A^<$ is indicated by the upper and the lower dash-dotted lines, 
the latter approaching zero rapidly for high momenta. The dotted line 
is the Hartree-Fock single particle spectrum. 
Density and temperature are the same as in Fig.~\ref{fig_sf}.
}
\end{figure}

\begin{figure}
\begin{center}
\epsfig{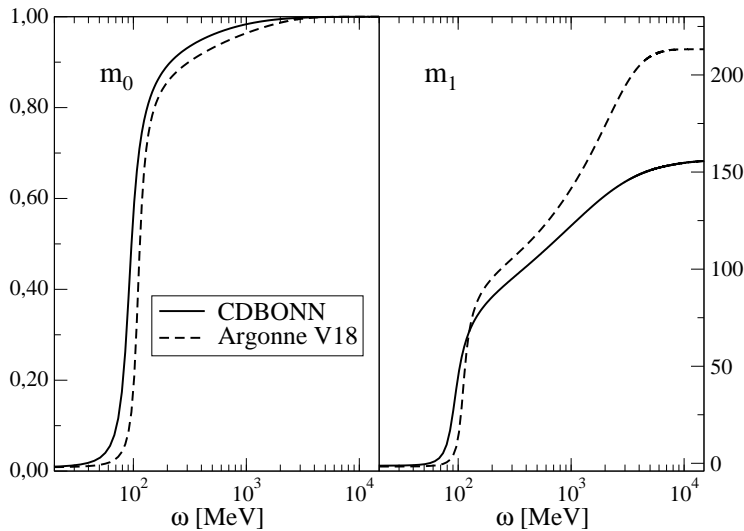}
\end{center}
\caption{\label{fig_sat}Saturation of the sum rules $m_0$ (right panel) 
and $m_1$ (left panel) for the CDBONN potential (solid line) and the
Argonne V18 potential (dashed line). The momentum is $k=500\,\mbox{MeV}$. 
Again, temperature and density are the same as in Fig.~\ref{fig_sf}.}
\end{figure}

The $m_0$ sum rule is fulfilled better than $0.1\%$ in the 
complete momentum range. 
Results for the $m_1$ sum rule are given in Fig~\ref{fig_m1} for a temperature 
of $T=10\,\mbox{MeV}$ and a density of $\rho=0.2\,\mbox{fm}^{-3}$. 
It is satisfied better than $1\%$. 
Both right hand side and left hand side are plotted, 
but the curves lie on top of each other and cannot be distinguished  
(solid lines). The lower dash-dotted line shows the $m_1$-contribution 
from $A^<$, which is always negative and goes to zero for high momenta, 
since there $A^<$ is strongly suppressed. The probability to remove a 
high-momentum particle from the system is simply very small.  
The upper dash-dotted line displays the contribution from $A^>$. Due to the 
short-range correlations, 
there is a high-energy tail present in the spectral function, and so
this contribution is already positive at low momenta, furthermore, it is
nearly constant in this range, 
reflecting the fact that the high energy strength distribution is 
momentum independent. As soon as the quasiparticle peak of the 
spectral function is located at energies greater than $\mu$, 
the $A^>$ contribution increases steadily, following the position of this 
peak. Both contributions add up to $m_1$. 
It is interesting to remind  the fact that for free particles, 
the sum rules are automatically fulfilled. In this case, 
$A^<$ and $A^>$ are delta peaks that are located at the same position 
and their strength adds up to one. Their relative strength is given by the 
ratio of the phase space factors $f(\epsilon(k))$ and 
$[1-f(\epsilon(k))]$, respectively, where $\epsilon(k)=\frac{k^2}{2m}$ 
in the free case.\\
The results in Fig.~\ref{fig_m1} shows that the sum rule 
$m_1$ is rather sensitive to the differences in the NN potentials.
The $m_1$ results for the CDBONN interaction is about 
$65\,\mbox{MeV}$
more attractive than the Argonne V18 result. 
A closer examination shows that this is predominantly due to 
the $A^>$ contribution, which is almost $50\,\mbox{MeV}$ more repulsive 
for the Argonne V18. This means that the Argonne potential 
produces more correlations in the sense that the  
strength that effects $m_1$ is redistributed to higher energies. \\
The dotted lines are the simple Hartree-Fock estimate of $m_1$ 
for the same temperature and density. For both potentials, 
the Hartree-Fock result makes up quite a good approximation to the sum rule.  
This result is interesting, since it permits a quantitative estimate of the
amount of correlations produced by any given NN potential without a 
sophisticated many-body calculation. \\
Fig.~\ref{fig_sat} reports the exhaustion of the sum 
rules $m_0$ (left panel) and $m_1$ (right panel) 
versus the upper integration limit $\omega$ for a 
momentum of $k=500\,\mbox{MeV}$. At this momentum, the quasiparticle 
peak is located around $100\,\mbox{MeV}$.
For both interactions that were considered, 
the main contribution to $m_0$, more than $80\%$, 
come from the quasiparticle peak of the spectral function.
In the region far above the peak, the CDBONN saturates considerably faster. 
In Table.~\ref{tab_sat}, the upper integration limits that have to be 
chosen to exhaust the sum rule to a given percentage are reported for 
$m_0$ and $m_1$ and compared for both potentials. In the case of 
$m_0$ and the stiffer Argonne V18, one must integrate almost twice as far as
for the softer CDBONN.
The saturation of the $m_1$ sum rule is different, 
because in this case, a somewhat higher energy region of the spectral function 
is probed.
In the right panel of Fig.~\ref{fig_sat}, one can observe that the 
quasiparticle peak contributes less than $50\%$ to $m_1$, and the 
high-energy tail becomes much more important, since it is weighted 
by a factor of $\omega$. While both potentials behave qualitatively 
similar up to an integration limit of about 
$700$ or $800\,\mbox{MeV}$, where $m_1$ is already exhausted by 
about 75\% for the CDBONN potential 
(cf.Tab.~\ref{tab_sat}), a large contribution
of about $40\%$ is still above this energy in the 
case of the Argonne V18. One can also note, that, 
to exhaust $m_1$ completely, one has to integrate up to higher 
energies in the case of the CDBONN.
However, these contributions to the spectral 
function above $\omega\approx 4000\,\mbox{MeV}$ are weak and
yield no further repulsion.

%$\mu=-25.5\,\mbox{MeV}$
%$\mu=-8.9\,\mbox{MeV}$

This work has been supported by the German-Spanish exchange program (DAAD,
Acciones Integradas Hispano-Alemanas).
We also would like to acknowledge financial support from the {\it Europ\"aische
Graduiertenkolleg T\"ubingen - Basel} (DFG - SNF) and the DGICYT (Spain) Project
No. BFM2002-01868 and from Generalitat de Catalunya Project No. 2001SGR00064.

\begin{table}
\begin{center}
\begin{tabular}{cccc}
\multicolumn{1}{c}{$k$ [MeV]} &
\multicolumn{1}{c}{below $\mu$ [\%]} &
\multicolumn{1}{c}{above $\mu$ [\%]} &
\multicolumn{1}{c}{$n(k)$} 
 \\ \hline\hline
230 & 98 & 2  & 0.706 \\
275 & 77 & 23 & 0.481 \\
320 & 33 & 67 & 0.191 \\
400 & 71 & 29 & 0.025 \\ 
500 & 95 & 5  & 0.006
\end{tabular}
\caption{\label{tab_dist}Strength distribution of $A^<$. The
numbers give the fraction of the integrated 
strength above and below the chemical
potential~$\mu$. The last column reports the occupation number of the
respective state.
The parameters are the same as in Fig.~\ref{fig_sf}.}
\end{center}
\end{table}

\begin{table}
\begin{center}
\begin{tabular}{ccccc}
\multicolumn{1}{c}{\% saturation} &
\multicolumn{1}{c}{$m_0$ CDB} &
\multicolumn{1}{c}{$m_0$ V18} &
\multicolumn{1}{c}{$m_1$ CDB} &
\multicolumn{1}{c}{$m_1$ V18}
 \\ \hline\hline
60 & -    & -    & 277  & 690  \\
75 & -    & -    & 790  & 1518 \\
90 & 215  & 311  & 2250 & 2860 \\
95 & 403  & 725  & 3756 & 3740 \\
99 & 1388 & 2277 & 8545 & 5720 \\
\end{tabular}
\caption{\label{tab_sat}Upper integration limits of the running integrals
that must be chosen to exhaust the sum rule $m_0$ and $m_1$ up to the 
fraction given in the first column. The parameters are the same as in 
Fig.~\ref{fig_sat}.}
\end{center}
\end{table}

\begin{thebibliography}{99}
\bibitem {bal1} M. Baldo, {\it Nuclear Methods and the Nuclear Equation of State,}
Int. Rev. of Nucl. Phys, Vol. 9 (World Scientific, Singapore, 1999).
\bibitem {her1} H. M\"uther and A. Polls, Prog. Part. Nucl. Phys. 
{\bf 45}, 243 (2000).
\bibitem{exp1} M. F. van Batenburg, 
Ph.D. thesis, University of Utrecht (2001).
\bibitem{her2} H. M\"uther, A. Polls and W. H. Dickhoff, Phys. Rev. {\bf C51},3040
(1995).
\bibitem{fan1} O. Benhar, A. Fabrocini and S. Fantoni,
Nucl. Phys. {\bf A 505}, 267 (1989). 
\bibitem{ram1} A. Ramos, A. Polls and W. H. Dickhoff, 
Nucl. Phys. {\bf A 503}, 1 (1989).
\bibitem{dick1} Y. Dewulf, W. H. Dickhoff, D. Van Neck,
E. R. Stoddard and M. Waroquier,
Phys. Rev. Let. {\bf 90}, 152501 (2003). 
\bibitem{dick2} W. H. Dickhoff and H. M\"uther, Rep. Prog. Phys. {\bf 55},
1947 (1992).
\bibitem{boz1} P. Bo\.zek, Phys. Rev. {\bf C 59}, 2619 (1999).
\bibitem{boz2} P. Bo\.zek, Phys. Rev. {\bf C 65}, 054306 (2002).
\bibitem{fri03} T. Frick and H. M\"uther, Phys. Rev. {\bf C 68}, 034310 (2003).
%\Journal{\PRC}{68}{034310}{2003}
\bibitem{pol94} A. Polls, A. Ramos, J. Ventura, S. Amari and
W. H. Dickhoff,
Phys. Rev. {\bf C 49}, 3050 (1994).
\bibitem{kraeft} W. D. Kraeft, D. Kremp, W. Ebeling and G. R\"opke, 
{\em Quantum Statistics of Charged Particle Systems} 
(Akademie-Verlag, Berlin, 1986).
\bibitem{kad62} L. P. Kadanoff and G. Baym,
{\em Quantum Statistical Mechanics} (Benjamin, New York, 1962).




\end{thebibliography}
\end{document}